# Team dynamics during the large-scale, engineered system

Christos Ellinas

*Abstract*— Coordinated collective action refers to the synchronized action of agents towards achieving a predefined set of goals. Such activity is at the core of a wide range of social challenges, from organizational dynamics to team performance. Focusing on the latter, a novel dataset that captures the planned efforts to deliver a large-scale, engineered system is introduced. In detail, this dataset is composed of a total of 271 unique individuals, responsible for the delivery of a total of 721 tasks spread across a period of 745 days. The focus of this analysis is on the collaboration network between individuals, captured by their co-assignment in the delivery of particular tasks, and their dynamical patterns. Results indicate that the delivery of some tasks depends on disproportionately large collaborations, making them intrinsically harder to manage compared to tasks which depend on small – or no – collaborations. Similarly, some tasks require a disproportionately diverse set of skills to be completed, further enhancing their intrinsic complexity. Shifting focus to the topology of the contribution network, an abrupt emergence, and subsequent contraction, of a single large cluster is observed. This phenomenon corresponds to the emergence of an increasingly large and cohesive team, and its subsequent decomposition. In addition, the evolution of this cluster tightly follows the number of active tasks, suggesting that large teams are a natural way to respond to increased workload. These findings provide new insight on the underlying team dynamics that govern coordinated collective activity in general, and in the context of project delivery specifically.

## I. INTRODUCTION

DELIVERING innovation relies on the collective activity of individuals who collaborate in a synchronized manner to generate a definitive, often pre-specified, outcome. This coordinated collective activity can be understood in the form of planned contributions, where individuals collaborate to complete sets of interconnected tasks, which are themselves spread across time. Handling these non-trivial patterns is a continuous challenge for both engineers and managers, with the former (engineers) being overwhelmed by the scale of information required to support collaboration, and the latter (managers) struggling to coordinate these collaborations. Yet, empirical studies that quantitatively analyze related social challenges are still lacking.

Equipped with large-scale datasets, smart algorithms and immense computational power, computational social science promises to fill this gap [1-3], with recent studies exploring a diverse set of related social challenges, ranging from predicting performance of individuals [4-6] to mapping contagion effects [7, 8]. Despite the explosive growth of the field, Watts [9] argues that "there has been surprisingly little progress on the 'big' questions that motivated the field in the first place – [including] questions concerning […] problem solving in complex organizations". This line of argument concludes to a need of shift in focus, from research merely striving to offer counter-intuitive, and often exotic [10], findings that may be of little practical relevance, to problem-driven, solution-oriented research [11-13].

For example, consider the recent stream of work that focuses on mapping the temporal dynamics of collective activity, reporting that such activity is typically 'busty', characterized by long periods of inactivity followed by short bursts of intense activity [14-16]. Despite the clear novelty of this finding and important theoretical consequences, their practical relevance is contextually restrained. For example, studies of 'bursty' activity appear to be limited to cases where individual action is unrestrained by activity of his/her peers (e.g. email communication, library visits etc.) i.e. it reflects collective but not coordinated collective action. Yet, a range of practical challenges rely stem from the 'coordinated' component, where the timing of an individual's actions is restrained by actions of other individuals. For example, when a group of individuals is assigned a set of activities, the order of these activities, and their interdependence, imposes strict limits to resulting actions. To further elaborate, consider a simple project composed of a linear chain of tasks: task $i$, where its completion unlocks task $j$, and task $j$'s completion unlocks the final task $k$; let us further assume that task $i$ is assigned to individual $a$, task $j$ to individual $b$, and task $k$ to individual $a$. In this case, individual $a$ is seen to be active for the duration of task $a$, and then remains necessarily inactive until task $j$ is completed, which will then unlock task $k$ and allow individual $a$ to become active again. If such chain continued *ad infinitum*, one would conclude that individual $a$'s activity is regular and periodic – yet this is merely the result of the underlying task structure and would certainly be different if tasks were to be structured in a different way. As such, we argue that understanding coordinated collective action necessitates an exposition of both tasks' and individuals' connectivity patterns, and how they evolve across time.

Christos Ellinas is with the University of Bristol, Senate House, Bristol BS8 1TH, UK (email: ce12183@bristol.ac.uk). This work was supported in part by an EPSRC Doctoral Prize fellowship (EP/N509619/1) and Thales UK.

With that in mind, the focus of this work is on team dynamics during the delivery of a large-scale engineering artefact. This endeavor is mapped by task and resources interdependencies, spread across time. In doing so, results indicate that some tasks require disproportionately large collaborations making them intrinsically harder to manage compared to tasks which require small – or no – collaborations. Similarly, some tasks require a disproportionately diverse set of skills to be completed, further enhancing their intrinsic complexity. Furthermore, tasks which depend on large collaborations connect with tasks that depend on small collaborations, suggesting the existence of communication bottlenecks. Shifting focus to the collaborations themselves, results demonstrate that the output of some pairwise collaborations is disproportionately important for the project to proceed, stressing the need to provide necessary conditions for these interaction to thrive. Finally, the abrupt emergence, and subsequent contraction, of a single project-wide cluster of interactions is reported.

## II. METHODS

### A. Data

The dataset captures the planned efforts to deliver a large-scale, engineered system in the defense domain. The dataset is composed of a total of $N^{\text{rsrcs}} = 271$ unique resources (or *individuals*), responsible for the delivery of $N^{\text{tsks}} = 443$ interdependent tasks, scheduled across $T = 745$ days (an additional 278 tasks exist, which are outsourced and hence, are omitted from this analysis). Every resource is classified in terms of one, potentially unique, specialization – we refer to characteristic as the resource's *role*.

This dataset naturally maps as a two-mode (or bipartite) network [17], which captures task-resource relationships – we refer to this construct as the *sociotechnical* network [18]. In this case, nodes correspond to tasks and resources, where links between the two capture the resource assignments for each task. In addition, the sociotechnical network can be projected to its one-mode counterparts, which reflect task-task relationships (*activity* network) and resource-resource relationships (*contribution* network), respectively. In the case of the activity network, nodes correspond to tasks, with weighted links between them capturing the number of resources that they share. In the case of the contribution network, nodes correspond to resources, with weighted links between them capturing the number of tasks that they both contribute to. The existence, and strength, of such link can serve as a proxy for collaboration, since a link between resource $i$ and $j$ means that they have to collaborate in order to deliver the tasks which they have in common.

Importantly, all three networks have an additional temporal dimension, which unlocks the ability to alternate between static and temporal aspects of the dataset. Static network properties can be explored by considering the aggregated network, where all links are accumulated across $T$ [19-21]. For example, the contribution network can be represented by an adjacency matrix $G^{\text{rsrc}} = \{a_{i,j}\} \in \mathbb{R}^{N^{\text{rsrcs}} \times N^{\text{rsrcs}}}$, where $a_{i,j}$ corresponds to the total number of times resource $i$ has collaborated with resource $j$ throughout $T$. Focusing on the temporal properties, the network can be viewed as a collection of snapshots, each reflecting the state of the network during a given time interval; in the case of the contribution network this can be represented by an adjacency matrix $G'^{\text{rsrc}} = \{a_{i,j,t}\} \in \mathbb{R}^{N^{\text{rsrcs}} \times N^{\text{rsrcs}} \times (T-\Delta t)}$, where $a_{i,j,t}$ is the *average* number of times resources $i$ and $j$ interact during the time interval $[t, t + \Delta t]$, where $\Delta t$ is a finite time-window; in this case, $\Delta t$ is set to 20 days (results are robust against varying $\Delta t$).

Lastly, note that various alternative definitions exist for activity networks e.g. [22-26] map activity networks based on functional dependencies i.e. task $i$ has an unweighted, directed link to task $j$ if the start of task $j$ depends on the completion of task $i$.

### B. Sociotechnical network

Every task has a number of unique resources assigned to it, given in the set $n^{\text{rsrc}}$, which is reflected by the blue node size in Fig. 1. Hence, a task with high $n^{\text{rsrc}}$ can be considered to be particularly demanding in terms of effort required for its completion, since it requires a large collaboration to be completed. With respect to the composition of a collaboration, the set $n^{\text{roles}}$ captures the number of unique roles that are needed for the delivery of each task. Hence, a task that requires a collaboration with a large (or small) $n^{\text{roles}}$ set suggests an increasingly interdisciplinary (or specialized) task.

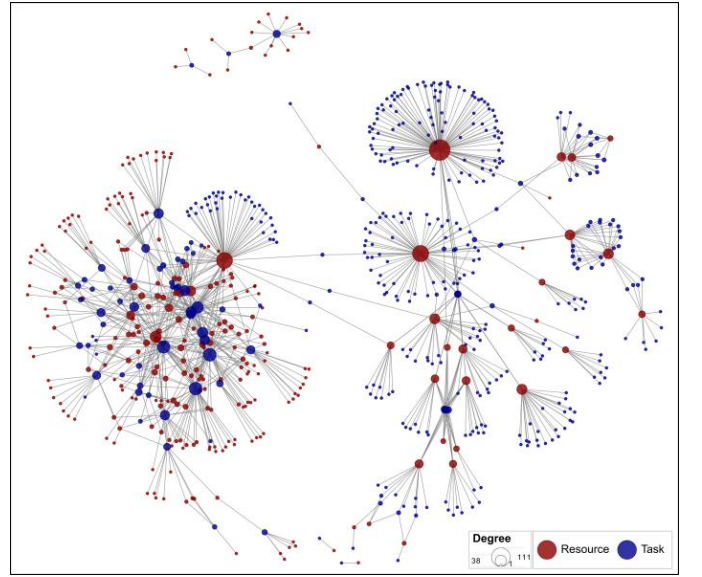

**Fig. 1**: Sociotechnical network, where tasks (blue) are linked with the resources (red) responsible for their delivery. Node size corresponds to the number of connections each task and/or resource has.

If two tasks are assigned to the same resource, these tasks are said to have 'shared ownership' [18], the extend of which is reflected by the size of $n^{\text{rsrcs}}$. For example, consider the case where resource $a$ contributes to the delivery of both task $i$ and $j$ – resource $a$ represents an individual who shares knowledge and responsibility for both task $i$ and $j$. As a result, task $i$ shares its 'ownership' with task $j$ due to their dependence on the same resource. With that in mind, consider the average 'shared ownership' of task's $i$ neighbours, referred to as 'mediated ownership' (defined as $n_i^{\text{med}} = \frac{1}{k_i}\sum_j^{k_i} n_j^{\text{rsrcs}}$, where $k_i$ corresponds to the total number of neighbors that task $i$ has). Tasks with a large $n^{\text{med}}$ are "likely to lie on organizational boundaries" and hence, serve as "loci for communication breakdowns", as argued in [18]. The overall relationship between $n^{\text{rsrcs}}$ and $n^{\text{med}}$ is also of interest, as it can provide additional insight on aspects related to the emergence of communication bottlenecks.

### C. Contribution network

Quantifying the importance of each *pair-wise* collaboration's output in terms of the overall project progression is of particular interest, as it can help identify critical links. One way for doing so it to identify the number of tasks each pair is responsible for, and evaluate the fraction of activity that these tasks occupy during that day. For example, if the outcome of a given pair-wise collaboration is the sole active task during that day, the collaboration's output importance is assigned a maximum value of 1; similarly, if the pair is responsible for 1 out of a total of 10 tasks active on that day, the collaboration's output importance is assigned a value of 0.1. Note that this measure focuses solely on the independent output of a pair, and does not consider the possibility of cascading failures [25, 27], where the failure of a pair to deliver its output can propagate across the project and affect the output of additional pairwise interactions.

Shifting focus from pair-wise to *group-wise* collaborations, we focus on the emergence of network clusters, where a cluster is defined as a group of nodes in which connections are dense, yet sparse between different groups [28]. In the context of the contribution network, a cluster represents a large collaboration in which all the resources within it are viewed as a distinct *team*. Hence, the evolution of teamwork throughout the project can be traced by considering how the average cluster size changes across time $T$. Doing so involves the following steps. First, a temporal version of the contribution network, represented by $G'^{\text{rsrc}}$, is generated. For each temporal slice, an implementation of Blondel, et al. [29]'s algorithm of the Newman-Girvan modularity measure [28, 30] is deployed, which classifies every active resource to an appropriate cluster – this process is reiterated 1,000 iterations to ensure convergence. Once each resource is assigned to an appropriate cluster, the average cluster size of each temporal slice is computed, with the entire process being reiterated for all $(T - \Delta t)$ slices. Note that the Newman-Girvan modularity measure is subjected to a resolution limitation – which can severely bias its accuracy [31] – unless the condition by Squartini and Garlaschelli [32] is satisfied [33], which is the case for this dataset.

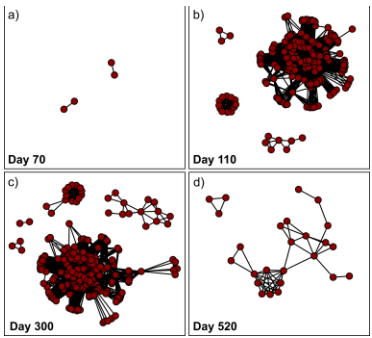

**Fig. 2**: a)-d) Topology of contribution network throughout the project, where nodes correspond to resources, and links reflect collaboration on the same tasks.

## III. RESULTS

### A. Role and resource assignments

The majority of tasks has a limited number of unique roles assigned to, reflecting their specialized nature. Conversely, a handful of tasks has an exceedingly high number of roles, reflecting their interdisciplinary nature. For example, the probability of observing a specialized task, with just 1 role, is over 0.7, yet some tasks rely heavily on a remarkably high number of resources, with 7 tasks being in the range $22 > n^{\text{roles}} \geq 11$ – see Fig. 3a. In other words, the project is largely composed of specialized tasks, complemented by tasks which are exceedingly interdisciplinary.

The distribution of $n^{\text{roles}}$ appears to be characterized by high variability, suggestive of a long tail. To determine whether this is the case, the empirical distribution is compared with an artificially constructed one, which decays exponentially fast by construction. Doing so involved generating an artificial ensemble of 1,000 normally distributed samples, with identical parameters (i.e. standard deviation and mean) which serve as a benchmark against the empirical distribution of $n^{\text{roles}}$ (for additional details on the methodology see [23, 34]). As expected, results indicate that the probability of observing specialized task is substantially higher in the empirical sample. Similarly, the extend of noted interdisciplinarity goes beyond the allowable range of the artificial sample, with the probability of observing it remaining relatively high within the empirical sample.

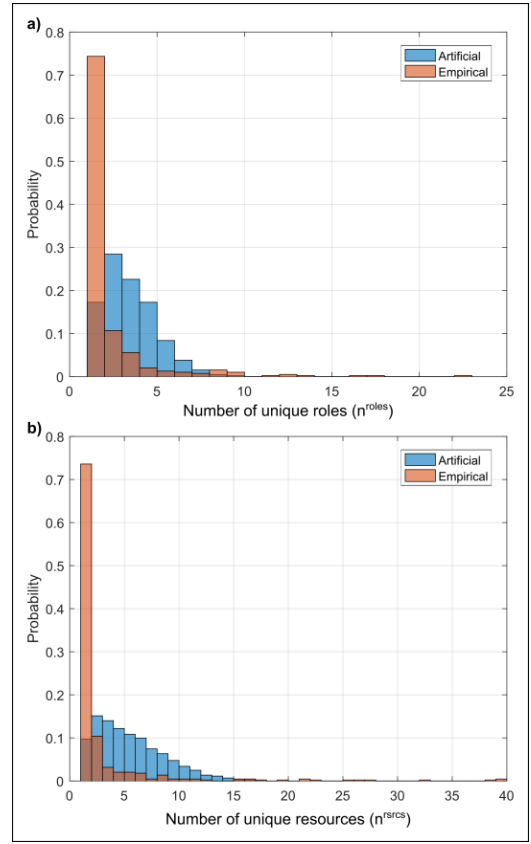

**Fig. 3**: a) Histogram of the number of roles $(n^{\text{roles}})$ any given task requires, within the empirical sample (orange) and it artificial counterpart (blue); b) similar to a), using the number of resources any given task requires $(n^{\text{rsrcs}})$.

Similarly, the majority of tasks requires a limited number of resources – indicative of small collaborations – with a few having an exceedingly high number, demonstrating the involvement of large collaborations (Fig. 3b). As expected, the empirical distribution of ($n^{\text{rsrcs}}$) is exceedingly different from its artificial counterpart, where the probability of observing a task requiring a collaboration of 2 is over 0.7, compared to a probability of 0.1 in the artificial dataset. In addition, there is a total of 16 tasks which require large collaborations ($40 > n^{\text{rsrcs}} \geq 15$), yet such extensive collaborations are not allowed under the artificial sample.

As expected, $n^{\text{rsrcs}}$ are $n^{\text{roles}}$ are highly correlated, where $n^{\text{rsrcs}}$ grows super-linearly with respect to increasing $n^{\text{roles}}$ (Spearman's $\rho = 0.967$; $p \leq 0.001$) – see Fig. 4a. This super-linear growth suggests that some roles are more important than others in delivering certain tasks i.e. a large portion of the required resources perform the same role.

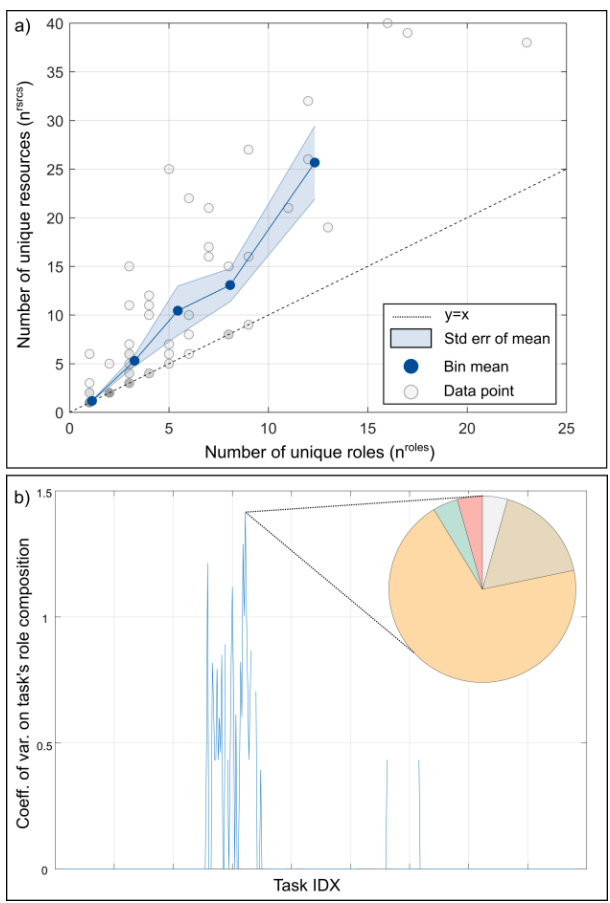

**Fig. 4**: a) Number of unique resources ($n_{\text{rsrcs}}$) as a function of the number of unique roles ($n_{\text{roles}}$), with bin averages (blue marker; bin edges determined using [35]) and a symmetry line (dotted line) representing linear growth; b) coefficient of variation for the number of resources with a given resource in any given task (higher score corresponds to less uniformity). The pie chart reflects the breakdown of Task IDX 341 in terms of the number of resources for each role assigned to it, indicating that a particular role (yellow) is dominant, with additional roles playing a significantly lower role.

Therefore, some tasks are non-uniform, in terms of their role requirements, and hence have the potential for disputes to arise between the assigned individuals due to cross-domain, interoperability-related challenges (e.g. terminology inconsistencies, variability if targets, different KPIs etc.).

Specifically, some tasks are exceedingly non-uniform in terms of their $n^{\text{roles}}$ composition, as quantified by the coefficient of variation ($cv$), where $cv_i = \frac{\sigma(n_i^{\text{roles}})}{\mu(n_i^{\text{roles}})}$ – see Fig. 4b. For example, Task IDX 341 ($n_{341}^{\text{rsrcs}} = 25$) is largely composed of resources that have the same role (70%), with the remaining resources being made up of 4 additional roles, each being a significantly smaller portion of the total number of resources assigned – see pie chart in Fig. 4b.

### B. Task ownership and importance

Tasks with 'high ownership' connect with tasks of relatively low 'shared ownership', on average, resulting in an overall negative trend, where $n^{\text{med}}$ decreases as $n^{\text{rsrc}}$ increases – see Fig. 5 (Spearman's $\rho = -0.497$; $p \leq 0.001$). As a result, these tasks "represent critical bottlenecks for expertise flow" [18] and hence, contribute to the challenging nature of the project. For example, consider task 130 ($n_{130}^{\text{rsrcs}} = 3$), which links to 3 tasks with identical ownership ($n_{130}^{\text{med}} = 1$). Hence, completing task 130 will require its assigned resources to coordinate with the resources responsible for delivering task 130's neighbors, which are few and proportionate in number, making communication/coordination relatively straightforward. However, consider task 7 ($n_7^{\text{rsrcs}} = 3$), which connects tasks with much higher 'shared ownership' ($n_7^{\text{med}} = 61.67$). In this case, resources responsible for the delivery of task 7 will have to coordinate with a disproportionately large number of resources, increasing the likelihood of communication/coordination challenges to emerge. More generally, the negative relationship between $n^{\text{med}}$ and $n^{\text{rsrc}}$ suggests that local coordination challenges (i.e. related to the delivery of a task) are inversely proportional to the broader coordination challenges (i.e. related to the delivery of a task's neighbors) that may arise.

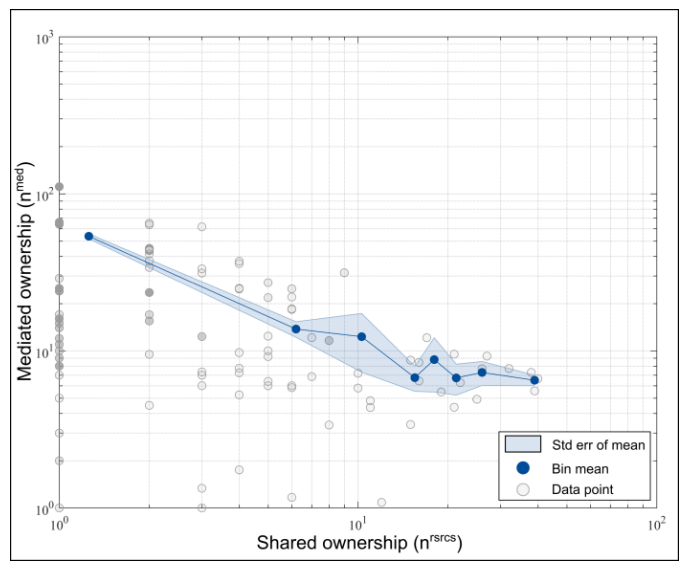

**Fig. 5**: 'Mediated ownership' ($n^{\text{med}}$) as a function of 'shared ownership' ($n^{\text{rsrcs}}$). Note the logarithmic nature of the axes.

### C. Pair-wise interactions

The output of the majority of pair-wise interactions is of limited importance, when considering the portion of daily progress that they control. In other words, the output of the majority of pair-wise interactions is a small fraction of the total number of active tasks on any given day. For example, the probability of encountering a pair-wise interaction that contributes just 1-1.5% of the total daily activity is more than 0.6 – see Fig. 6. At the same time, a handful of tasks exist which have a disproportionally important role in the daily progression of the project, with some interaction reaching a maximum contribution of up to 25% of the total daily activity. Clearly, these pair-wise interactions are of critical importance to the overall progression, and hence should be clearly identified and nourished.

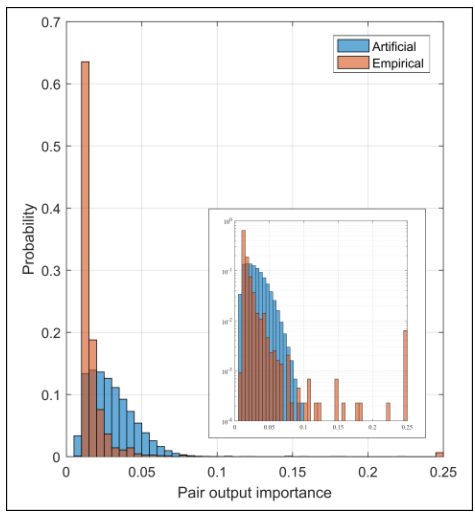

**Fig. 6**: Empirical distribution of each pair's output importance (red), compared to its normally distributed counterpart (blue). Inset is the same using log scale on the y-axis.

More generally, the distribution of the importance of all pair-wise contributions is heavily skewed. Compared to its normally-distributed, artificial counterpart, low values are significantly overrepresented, resulting to the majority of interactions having little importance in terms of overall project daily progress. At the same time, the empirical distribution sustains a much larger range of values to be observed compared to the artificial sample, resulting in disproportionally important pair-wise interactions.

*D. Group-wise interactions*

The project undertakes an abrupt expansion, and subsequent contraction, of a large, project-wide collaboration throughout its duration, with additional smaller collaborations interacting with it in a dynamic way. This effect can be efficiently captured using a proximity timeline (Fig. 7), where each line corresponds to a resource within the contribution network, and their proximity reflects the number links between them (i.e. geodesic distance) [36].

The abrupt emergence of this large-scale collaboration is evident by a sharp transition from individual strands to one prominent strand, at $t = 100$, which subsequently breaks up at $t = 510$. This phenomenon corresponds to the emergence of an increasingly large and cohesive team following project initiation, and its subsequent decomposition, as the project enters the closing phase. During this period, several additional strands emerge, some of which are initially independent of the main backbone, followed by intermittent interaction periods with the larger strand, whilst other remain independent through the entire project (Fig. 7; blue strand). After this time ($t > 510$), the large backbone breaks down, with the contribution network sustaining few and relatively small collaborations until the majority of resources defaults to the original work mode (i.e. isolated).

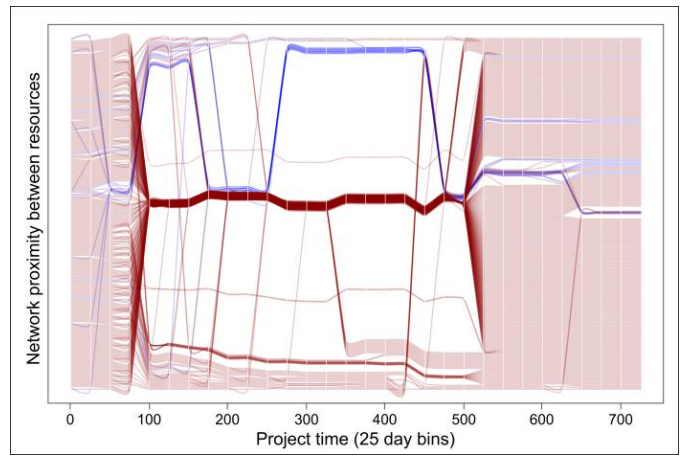

**Fig. 7**: Proximity timeline of the contribution network, where each link corresponds to a resource, and the relative distance between them reflects the number of links between them i.e. dense strands correspond to clusters; distinct strands correspond to isolated resources. Blue strand is an example team which intermittently interacts with the core team (dense red strand)

Interestingly, a strong correlation between the average cluster size and the number of active tasks exists, suggesting that large teams are a natural way to respond to increased workload (Fig. 8). However, this correlation may simply be a residual of having many tasks running in parallel, which in turn fuels this cluster due to the accumulation of resources that have to be active during that time. As such, a question arises of whether the emergence of this large cluster is the result of purposeful actions (e.g. planning an increased amount of collaboration in order to handle the increasingly high daily workload) or merely random (i.e. due to the increasingly high number of active tasks, and the consequent high number of active resources assigned to them). Untangling the two requires the construction of a null model that simulates the latter process, with its output serving as a benchmark for comparing the empirical effect noted in Fig. 7, and the output of the null model.

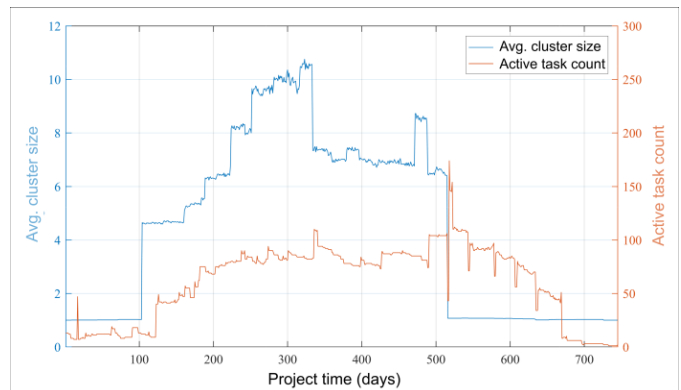

**Fig. 8**: Temporal evolution of the average cluster size within the contribution network (blue) and the number of active tasks (red).

## IV. DISCUSSION

Coordinated collective action is at the core of a range of challenges, including that of understanding teamwork dynamics, as in the case of delivering a definitive output. In this particular example, readily-available, empirical data has been used to explore the characteristics and underlying dynamics of a coordinated endeavor to deliver a complex engineering artefact. By viewing this endeavor as a sociotechnical network –and its composing contribution network – the focus of this work was on evaluating the heterogeneity of key variables, which have recently been argued to control the performance of such endeavors [23, 37, 38]. In doing so, different levels of task heterogeneity are identified, with the majority of tasks being delivered by small – or no – collaborations, with each collaboration being largely composed of identical roles. Despite this seemingly homogeneity, a handful of tasks stretch the range of both, requiring exceedingly large, and interdisciplinary, collaborations (Fig. 3). Assuming that communication/coordination frameworks are set based on average characteristics, one could reasonably postulate that such extreme deviations would play a key role in undermining these frameworks – therefore identifying them and appropriately managing them is important. At the same time, some of these interdisciplinary collaborations are dominated by specific roles, potentially triggering additional challenges related to cross-discipline interoperability (Fig. 4), where team members form cliques between members of same roles (e.g. due to co-location of people with the same role) stressing communication channels further.

The degree of this heterogeneity extends further once interconnectivity is take into account, with several tasks having increased levels of 'mediated ownership' (Fig. 5). The existence of such task poses a risk for communication bottlenecks to emerge, since the respective individuals must successfully manage and mediate information flow to a disproportionally large number of individuals in order to maintain a smooth transition between the completion of interdependent tasks. Interestingly, the existence of an inverse relationship between 'mediated ownership' and 'shared ownership' stresses the importance of interconnectivity: if one was to consider tasks independently, an assignment of few individuals would suggest that limited coordination/communication challenges will arise during its delivery. Alas, these tasks appear to be, on average, more likely to act as intermediates to task with a large number of individuals, stressing the need for increased preparedness in order to cope with the increased levels of information flow that will need to be managed.

Considering the temporal dynamics of the contribution network, a sharp transition is noted, from a period in which individual work dominates, to a period where large-scale collaborations dominate. In addition, this behavior is in step with the number of active tasks, suggesting that the teamwork is a natural way for tackling increased workload (Fig. 8). This behavior sheds light into the imposed coordination dynamics, where the vast majority of collaborations are planned mid-project. In conjunction with the fact that the dominance of large-scale clusters can facilitate global spreading events (through intracommunity spreading [39]), periods of heighted collaboration can be critical in terms of their potential of triggering cascading failures [40], diffusing undesirable behavioral traits [41] etc. More generally, the extend of this period in which large-scale collaborations dominate could provide a proxy for the susceptibility of coordinated collective action to spreading phenomena, which in turn can damage their effectiveness/efficiency across multiple facets. Further work in exploring this line of enquiry could clarify the precise effect of this phenomenon, and how it perform with respect to other related phenomena (e.g. temporal correlation between activities [19] etc.).

## V. CONCLUSION

This works introduces the challenge of understanding coordinated collective action in general, and in the context of team dynamics specifically. Focusing on planned efforts to deliver a large-scale, engineered system, results demonstrate that: (a) majority of tasks require no collaborations, with some requiring exceedingly large collaborations, (b) same as (a) but with respect to the number of unique roles required, (c) tasks that require small collaborations (low 'shared ownership') are connected with tasks that require large collaborations (high 'mediated ownership'), and (d) a large-scale collaboration abruptly emerges relatively early in the project, and subsequently abruptly decomposes as the project approaches termination. Finally, this works argues that the challenge presented – coordinated collective action – and the contextual example of team activity are general enough to undermine a range of social challenges, and sufficiently practical to provide for solution-oriented, future research.